\documentclass{aa}  
\usepackage{graphicx}
\usepackage{txfonts}
\begin{document}

   \title{
On the co-existence of chemically peculiar Bp stars, slowly pulsating B stars and constant B stars in the same part of the H-R diagram
}

   \author{M. Briquet\inst{1}\thanks{Postdoctoral Fellow of the Fund for Scientific Research, Flanders}
          \and
          S. Hubrig\inst{2}
          \and
          P. De Cat\inst{3} 
          \and
          C. Aerts\inst{1,4}
          \and
          P. North\inst{5}
          \and
          M. Sch\"oller\inst{2}
          }

   \offprints{M.Briquet}

   \institute{  Instituut voor Sterrenkunde, Katholieke Universiteit Leuven, Celestijnenlaan 200 D, B-3001 Leuven, Belgium \\
              \email{maryline@ster.kuleuven.be}
         \and 
               European Southern Observatory, Casilla 19001, Santiago 19, Chile        
         \and             
              Koninklijke Sterrenwacht van Belgi\"e, Ringlaan 3, B-1180 Brussel, Belgium
         \and            
              Department of Astrophysics, University of Nijmegen, PO Box 9010, 6500 GL Nijmegen, the Netherlands
         \and         
              Laboratoire d'astrophysique, Ecole Polytechnique F\'ed\'erale de Lausanne (EPFL), Observatoire, CH-1290 Sauverny, Switzerland       
             }

   \date{Received; accepted}

% \abstract{}{}{}{}{} 
% 5 {} token are mandatory
 
  \abstract
  % context heading (optional)
  % {} leave it empty if necessary  
   {}
  % aims heading (mandatory)
   {
In order to better model massive B-type stars,
we need to understand the physical processes taking place 
in slowly pulsating B (SPB) stars, chemically peculiar Bp stars,
and non-pulsating normal B stars co-existing in the same part of the H-R diagram.
  }
  % methods heading (mandatory)
   {
We carry out a comparative study between samples
of confirmed and well-studied SPB stars
and a sample of well-studied Bp stars
with known periods and magnetic field strengths.
We determine their evolutionary state using
accurate HIPPARCOS parallaxes and Geneva photometry.
We discuss the occurrence and strengths of magnetic fields
as well as the occurrence of stellar pulsation among both groups.
Further, we make a comparison of Geneva photometric variability
for both kinds of stars.
  }
  % results heading (mandatory)
   {
The group of Bp stars is significantly younger than the group of SPB stars.
Longitudinal magnetic fields in SPB stars are weaker than those of Bp stars,
suggesting that the magnetic field strength is an important factor
for B type stars to become chemically peculiar.
The strongest magnetic fields appear in young Bp stars,
indicating a magnetic field decay in stars at advanced ages.
Rotation periods of Bp and pulsation periods of SPB stars are of the same order
and the behaviour of Geneva photometric variability
of some Bp stars cannot be distinguished from the variability of SPB stars,
illustrating the difficulty to interpret the observed variability
of the order of days for B-type stars.
We consider the possibility that pulsation could be responsible for the variability among chemically peculiar stars.
In particular, we show that a non-linear pulsation model
is not excluded by photometry for the Bp star HD\,175362.  
  }
  % conclusions heading (optional), leave it empty if necessary 
   {}

   \titlerunning{On the co-existence of Bp stars, SPB stars and constant B stars in the same part of the H-R diagram}
   \maketitle

%
%________________________________________________________________

\section{Introduction}

\begin{table*}
\caption{
Fundamental parameters of Bp stars, SPB stars and normal B stars.
We give, in order,
HD number,
rotation period or main pulsation period for Bp stars and SPB stars
respectively,
effective longitudinal magnetic field,
absolute visual magnitude,
mass,
effective temperature,
luminosity,
surface gravity,
radius,
distance
and relative uncertainty of parallax.
 }
\label{Bp_SPB_B}
\centering
\begin{tabular}{c c c c c c c c c c c} 
\hline
HD & $P$ [d] & $\left<B_l\right>$ [G] & M$_v$ & $M / M_\odot$ & $\log(T_{\rm eff})$ & $\log(L / L_\odot)$ & $\log g$ & $R / R_\odot$ & d [pc] & $\sigma(\pi) / \pi$\\
\hline
\multicolumn{11}{c}{Bp stars}\\
\hline
    5737&21.6454 & 324&$-$2.27 & 4.976 $\pm$ 0.335 & 4.121 $\pm$ 0.013 & 3.068 $\pm$ 0.156 &  3.50 $\pm$ 0.14 &  6.54 $\pm$ 1.24 & 206&  0.173 \\
   12767&1.892 &  242&$-$0.54 & 3.643 $\pm$ 0.152 & 4.111 $\pm$ 0.013 & 2.369 $\pm$ 0.085 &  4.03 $\pm$  0.09 &  3.06 $\pm$ 0.35 & 111 & 0.086  \\
   19832 &0.727893 &315 &0.34 & 3.142 $\pm$ 0.144 & 4.095 $\pm$ 0.013 & 2.008 $\pm$ 0.096 &  4.26 $\pm$ 0.10 &  2.17 $\pm$ 0.27 & 114 & 0.100  \\
   21699&2.4765 &828 &$-$1.06 & 4.314 $\pm$ 0.235 & 4.159 $\pm$ 0.013 & 2.660 $\pm$ 0.118 &  4.00 $\pm$ 0.11 &  3.44 $\pm$  0.51 & 180 & 0.127  \\
   22470&0.6785 &733 &$-$0.67 & 3.736 $\pm$ 0.179 & 4.115 $\pm$ 0.013 & 2.424 $\pm$ 0.111 &  4.00 $\pm$ 0.11 &  3.20 $\pm$  0.45 & 145 & 0.119 \\ 
   24155&2.53465  &1034 &0.30 & 3.353 $\pm$ 0.176 & 4.132 $\pm$ 0.013 & 2.059 $\pm$ 0.112 &  4.38 $\pm$ 0.11 &  1.95 $\pm$  0.28 & 136 & 0.121 \\ 
   25823 &4.65853 &668 &$-$0.48 & 3.615 $\pm$ 0.234 & 4.112 $\pm$ 0.026 & 2.343 $\pm$ 0.119 &  4.05 $\pm$ 0.15 &  2.95 $\pm$ 0.54 & 152 & 0.129  \\
   28843&1.373813 & 344 &0.05 & 3.555 $\pm$ 0.173 & 4.143 $\pm$ 0.013 & 2.179 $\pm$ 0.103 &  4.33 $\pm$ 0.10 &  2.13 $\pm$ 0.28 & 131&  0.109  \\
   34452&2.4687 &527 &$-$0.32 & 3.754 $\pm$ 0.212 & 4.158 $\pm$ 0.026 & 2.270 $\pm$  0.093 &  4.33 $\pm$ 0.14 &  2.20 $\pm$ 0.35 & 137&  0.096  \\
   49333&2.18010 &618 &$-$0.56 & 4.289 $\pm$ 0.262 & 4.182 $\pm$ 0.013 & 2.538 $\pm$ 0.134 &  4.21 $\pm$ 0.12 &  2.68 $\pm$ 0.44 & 205&  0.148  \\
   64740&1.33026 &587 &$-$2.15 & 8.740 $\pm$ 0.407 & 4.353 $\pm$ 0.013 & 3.761 $\pm$ 0.107 &  3.98 $\pm$ 0.10 &  4.99 $\pm$ 0.69 & 221&  0.115  \\
   73340&2.66753 &1644 &$-$0.11 & 3.667 $\pm$ 0.130 & 4.145 $\pm$ 0.013 & 2.253 $\pm$  0.068 &  4.28 $\pm$  0.08 &  2.30 $\pm$ 0.23 & 143&  0.063  \\
   92664 &1.67315 &803 &$-$0.32 & 3.863 $\pm$ 0.147 & 4.154 $\pm$ 0.013 & 2.357 $\pm$  0.075 &  4.24 $\pm$  0.09 &  2.47 $\pm$ 0.26 & 143&  0.073  \\
 125823 &8.8171&328& $-$1.08& 5.882 $\pm$ 0.260& 4.265 $\pm$ 0.012& 3.054 $\pm$ 0.094&  4.16 $\pm$  0.09&  3.32 $\pm$ 0.40& 128& 0.098 \\
 133652&2.3040&1116 &0.74 & 3.050 $\pm$ 0.132 & 4.113 $\pm$ 0.013 & 1.863 $\pm$ 0.089 &  4.46 $\pm$  0.09 &  1.69 $\pm$ 0.20 &  96 & 0.092  \\ 
 137509&4.4912 &1062 &$-$0.29 & 3.367 $\pm$ 0.203 & 4.076 $\pm$ 0.022 & 2.268 $\pm$ 0.138 &  3.95 $\pm$ 0.14 &  3.21 $\pm$ 0.60 & 249 & 0.152  \\
 142301&1.45955 &2104 &$-$0.28 & 4.243 $\pm$ 0.289 & 4.193 $\pm$ 0.013 & 2.470 $\pm$ 0.151 &  4.32 $\pm$0.14 &  2.36 $\pm$ 0.43 & 140 & 0.168  \\
 142990&0.9791 &1304 &$-$0.80 & 4.902 $\pm$ 0.260 & 4.217 $\pm$ 0.013 & 2.765 $\pm$ 0.114 &  4.18 $\pm$ 0.11 &  2.97 $\pm$ 0.43 & 150 & 0.123  \\
 144334&1.49497 &783 &$-$0.31 & 4.085 $\pm$ 0.256 & 4.167 $\pm$ 0.024 & 2.465 $\pm$ 0.118 &  4.20 $\pm$ 0.15 &  2.65 $\pm$ 0.46 & 149 & 0.128  \\
 147010&3.920676 &4032 &0.59 & 3.137 $\pm$ 0.180 & 4.117 $\pm$ 0.013 & 1.931 $\pm$ 0.125 &  4.42 $\pm$  0.12 &  1.80 $\pm$  0.28 & 143 & 0.136  \\
 151965&1.60841 &2603 &$-$0.10 & 3.736 $\pm$ 0.245 & 4.154 $\pm$ 0.013 & 2.272 $\pm$ 0.145 &  4.31 $\pm$ 0.13 &  2.25 $\pm$ 0.40 & 181 & 0.161  \\
 168733&14.78437 &815 &$-$1.16 & 4.015 $\pm$ 0.230 & 4.108 $\pm$ 0.020 & 2.614 $\pm$ 0.131 &  3.81 $\pm$  0.14 &  4.12 $\pm$ 0.73 & 190 & 0.144  \\
 175362&3.67375 &3569 &$-$0.38 & 3.986 $\pm$ 0.267 & 4.164 $\pm$ 0.030 & 2.404 $\pm$ 0.114 &  4.24 $\pm$ 0.17 &  2.50 $\pm$ 0.47 & 130 & 0.123  \\
 196178&1.91645 &1069 &$-$0.13 & 3.542 $\pm$ 0.162 & 4.126 $\pm$ 0.013 & 2.227 $\pm$  0.096 &  4.22 $\pm$  0.10 &  2.43 $\pm$ 0.30 & 147 & 0.100  \\
\hline
\multicolumn{11}{c}{SPB stars}\\
\hline
  3360&1.5625 & $-$& $-$2.78& 8.582 $\pm$ 0.388& 4.324 $\pm$  0.004& 3.827 $\pm$  0.105&  3.79 $\pm$  0.09&  6.15 $\pm$ 0.75& 183& 0.112 \\
  3379&0.54937 & 182&$-$1.38& 5.524 $\pm$ 0.398& 4.237 $\pm$  0.004& 3.024 $\pm$  0.167&  4.06 $\pm$ 0.14&  3.64 $\pm$ 0.70& 262& 0.186 \\
 21071&0.84145 & $-$&$-$0.51& 3.999 $\pm$ 0.230& 4.157 $\pm$  0.003& 2.453 $\pm$  0.133&  4.17 $\pm$ 0.11&  2.74 $\pm$ 0.42& 185& 0.146 \\
 24587&0.86438 & $-$&$-$0.69& 3.969 $\pm$ 0.149& 4.141 $\pm$  0.002& 2.496 $\pm$  0.087&  4.06 $\pm$  0.07&  3.09 $\pm$ 0.31& 118& 0.089 \\
 26326&1.87336 & $-$&$-$1.39& 4.820 $\pm$ 0.324& 4.182 $\pm$  0.003& 2.871 $\pm$  0.156&  3.93 $\pm$ 0.13&  3.94 $\pm$ 0.71& 223& 0.174 \\
 27026& 0.61387 & $-$& 0.70& 2.936 $\pm$ 0.118& 4.082 $\pm$  0.002& 1.869 $\pm$  0.093&  4.32 $\pm$  0.08&  1.97 $\pm$ 0.21& 119& 0.097 \\
 27396&2.16826 & $-$&$-$1.42& 4.879 $\pm$ 0.222& 4.185 $\pm$  0.003& 2.892 $\pm$  0.105&  3.93 $\pm$  0.09&  3.98 $\pm$ 0.49& 142& 0.112 \\
 28114&2.04842 & $-$&$-$0.81& 4.228 $\pm$ 0.305& 4.163 $\pm$  0.003& 2.589 $\pm$  0.167&  4.08 $\pm$ 0.14&  3.11 $\pm$ 0.60& 183& 0.187 \\
 34798&1.27632 & $-$&$-$0.63& 4.458 $\pm$ 0.337& 4.193 $\pm$  0.003& 2.600 $\pm$  0.175&  4.21 $\pm$ 0.14&  2.74 $\pm$ 0.55& 243& 0.197 \\
 53921&1.65180 & 185&$-$0.27& 3.714 $\pm$ 0.129& 4.137 $\pm$  0.002& 2.322 $\pm$  0.080&  4.18 $\pm$  0.07&  2.58 $\pm$ 0.24& 148& 0.080 \\
 74195&2.81889 & 200&$-$2.32& 6.042 $\pm$ 0.204& 4.208 $\pm$  0.003& 3.320 $\pm$  0.078&  3.69 $\pm$  0.07&  5.85 $\pm$ 0.53& 152& 0.077 \\
 74560&1.55106 & 146&$-$0.97& 4.863 $\pm$ 0.152& 4.210 $\pm$  0.004& 2.783 $\pm$  0.071&  4.13 $\pm$  0.06&  3.13 $\pm$ 0.26& 147& 0.068 \\
 92287&4.65549 & $-$&$-$2.18& 5.978 $\pm$ 0.459& 4.215 $\pm$  0.004& 3.283 $\pm$  0.178&  3.74 $\pm$ 0.15&  5.43 $\pm$ 1.12& 392& 0.200 \\
123515&1.45926 & $-$&$-$0.03& 3.235 $\pm$ 0.151& 4.079 $\pm$  0.002& 2.154 $\pm$  0.108&  4.06 $\pm$  0.09&  2.77 $\pm$ 0.35& 168& 0.116 \\
138764&1.25881 & $-$&$-$0.22& 3.777 $\pm$ 0.147& 4.148 $\pm$  0.003& 2.321 $\pm$  0.090&  4.24 $\pm$  0.07&  2.45 $\pm$ 0.25& 108& 0.092 \\
140873&0.868432 & 231&$-$0.31& 3.788 $\pm$ 0.138& 4.144 $\pm$  0.003& 2.349 $\pm$  0.084&  4.19 $\pm$ 0.07&  2.58 $\pm$ 0.25& 125& 0.085 \\
147394&1.24958 & $-$&$-$1.06& 4.403 $\pm$ 0.115& 4.165 $\pm$  0.003& 2.692 $\pm$  0.060&  4.00 $\pm$  0.05&  3.46 $\pm$ 0.24&  96& 0.051 \\
160762& 4.02739 & $-$&$-$2.09& 6.136 $\pm$ 0.222& 4.234 $\pm$  0.004& 3.299 $\pm$  0.084&  3.82 $\pm$  0.07&  5.06 $\pm$ 0.50& 152& 0.085 \\
169820&0.47057 & 147& 0.87& 2.789 $\pm$ 0.111& 4.071 $\pm$  0.002& 1.770 $\pm$  0.092&  4.35 $\pm$  0.08&  1.85 $\pm$ 0.20& 115& 0.095 \\
181558&1.23793 & 201&$-$0.58& 4.130 $\pm$ 0.274& 4.166 $\pm$  0.003& 2.504 $\pm$  0.154&  4.17 $\pm$ 0.13&  2.77 $\pm$ 0.49& 208& 0.171 \\
182255&1.26216 & $-$&$-$0.36& 3.860 $\pm$ 0.139& 4.149 $\pm$  0.003& 2.379 $\pm$  0.083&  4.19 $\pm$  0.07&  2.60 $\pm$ 0.25& 123& 0.084 \\
206540&1.38887 & $-$&$-$0.71& 4.014 $\pm$ 0.269& 4.145 $\pm$  0.003& 2.512 $\pm$  0.155&  4.06 $\pm$ 0.13&  3.08 $\pm$ 0.55& 214& 0.173 \\
208057&1.24732 & 133&$-$0.99& 5.007 $\pm$ 0.225& 4.220 $\pm$  0.004& 2.817 $\pm$  0.103&  4.15 $\pm$  0.09&  3.12 $\pm$ 0.38& 157& 0.110 \\
215573&1.83857 & 174&$-$0.42& 3.844 $\pm$ 0.114& 4.144 $\pm$  0.003& 2.393 $\pm$  0.068&  4.16 $\pm$  0.06&  2.71 $\pm$ 0.22& 136& 0.064 \\ 
\hline
\multicolumn{11}{c}{normal B stars}\\
\hline
134837& $-$ & $-$ & 0.75& 2.997 $\pm$ 0.116& 4.097 $\pm$  0.002& 1.877 $\pm$ 0.089&  4.38 $\pm$ 0.07&  1.86 $\pm$ 0.19& 111& 0.092 \\
142378& $-$ & $-$ &  $-$0.99& 4.832 $\pm$ 0.350& 4.206 $\pm$  0.004& 2.782 $\pm$ 0.167&  4.12 $\pm$ 0.14&  3.18 $\pm$ 0.62& 191& 0.187 \\
164245& $-$ & $-$ &  $-$0.81& 3.870 $\pm$ 0.271& 4.112 $\pm$  0.002& 2.520 $\pm$ 0.162&  3.91 $\pm$ 0.13&  3.63 $\pm$ 0.68& 221& 0.181 \\
205265& $-$ & $-$ &  $-$0.36& 3.634 $\pm$ 0.282& 4.117 $\pm$  0.002& 2.342 $\pm$ 0.180&  4.08 $\pm$ 0.15&  2.89 $\pm$ 0.60& 213& 0.203 \\
212986& $-$ & $-$ &  $-$0.58& 3.961 $\pm$ 0.221& 4.148 $\pm$  0.003& 2.463 $\pm$ 0.129&  4.12 $\pm$ 0.11&  2.88 $\pm$ 0.43& 249& 0.142 \\
\hline
\end{tabular}\\

\end{table*}

Different types of B stars co-exist at the same position of the H-R diagram, which coincides with the instability strip 
of slowly pulsating B stars (SPB stars). SPB stars are very promising targets for asteroseismic studies because 
these B-type stars with masses between 3 and 9\,M$_\odot$ pulsate in many high-order gravity-modes. This makes it 
possible to probe very deep layers in the stellar interior of this kind of stars. Beside SPB stars, we find chemically 
peculiar Bp stars, which show abnormal abundances of certain chemical elements in their atmosphere. They possess 
variable magnetic fields and show light and line-profile variations which are interpreted within the framework 
of the oblique rotator model. As the star rotates, we observe the magnetic field and inhomogeneous surface abundance 
distributions from various aspects, resulting in the observed variability. In addition, there are also non-pulsating 
normal B stars at the same position of the H-R diagram. 

With the aim to improve stellar structure models of high mass stars we need to explain the co-existence of these three 
groups of stars in the SPB instability strip. Several physical processes have been put forward, but a clear scenario 
is not yet established. For instance, a slow rotation rate is assumed a necessary condition for a star to become 
chemically peculiar, but it does not seem to be a sufficient condition since both the groups of SPB stars and Bp stars 
consist of slow rotators. A difference of metallicity between SPB stars and normal B stars could be invoked to 
explain that only some main-sequence B-type stars reach observable pulsation amplitudes. However, the recent study of 
Niemczura\ (\cite{niemczura}) showed that SPB stars do not differ from the normal B-type stars as far as the 
metallicity is concerned. Recently, we presented the results of a magnetic survey of a sample of 26 SPB stars 
with FORS\,1 at the VLT.  A weak
mean longitudinal magnetic field of the order of a few hundred Gauss has been detected
in 13 SPB stars (Hubrig et al.\ \cite{hubrig3}). All magnetic SPB stars for which we gathered several 
magnetic field measurements show a field that varies in time.  However, we were not able to find any relation between 
the fundamental parameters of this group of stars and the presence of magnetic fields in their atmospheres.

To search for the connections and differences that could exist between the different types of stars, we have carried out 
a comparative study between a sample of well-studied Bp stars with known periods and magnetic field strengths, 
a sample of confirmed and well-studied B-type pulsators and a sample of normal B stars. The selection of our star 
samples is described in Sect.\,2. In Sect.\,3 we determine the fundamental stellar parameters. In Sect.\,4 we 
compare the three groups, by focussing on stellar evolutionary state, stellar rotation and stellar magnetic field strengths. In Sect.\,5 we perform a comparison of Geneva photometric variability of Bp and SPB stars.
We end with a discussion in Sect.\,6.

\section{Selection of star samples}
We selected our sample of magnetic Bp stars from the recent catalogues of Bychkov et al.\ (\cite{bychkov1}) and 
Hubrig et al.\ (\cite{hubrig2}). We considered stars with masses between 3 and 9\,M$_\odot$, for which the periods 
and magnetic field strengths are known. Note that we did not include Bp stars with the PGa and HgMn peculiarity as 
the presence of very weak magnetic fields has been confirmed only in a small sample of these stars. Furthermore, 
the structure of 
the detected magnetic field in these stars should be sufficiently tangled as it does not produce a strong net 
observable circular polarization signature (Hubrig et al.\ \cite{hubrig2}).
Our selected sample of Bp stars consists of He-weak stars and Si stars with $T_{\rm eff}$ in the same range as 
the studied SPB stars. The hottest star is He-rich.

The number of candidate SPB stars has increased from 12 to more than 80 thanks to the Hipparcos 
mission (Waelkens et al.\ \cite{waelkens}). The way of classification used by the latter authors does not allow to 
discriminate chemically peculiar variables falling in the SPB domain from real SPB stars. Therefore, it may be possible
that some candidate SPB stars are actually Bp stars, as was discovered for four stars by means of a detailed 
spectroscopic study (Briquet et al.\ \cite{briquet}). For this reason, only confirmed SPB stars were included 
in our SPB star sample. The list of stars was retrieved from De Cat (\cite{decat1}). The normal B stars were 
selected as standard B-type stars in the photometric system of Geneva.

\begin{figure*}
\centering
\includegraphics[angle=-90,width=13cm]{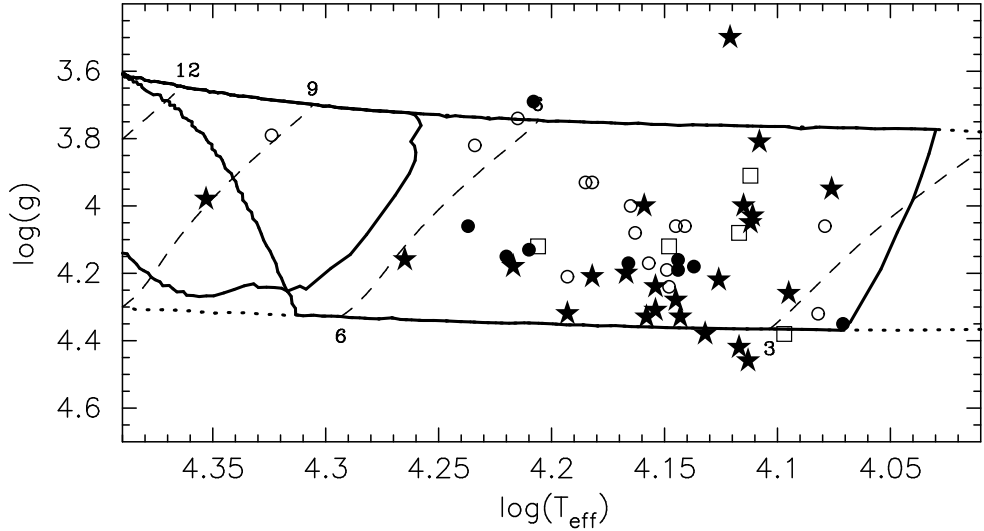}
\caption{The position of the stars in our samples in a ($\log T_{\rm eff},\log g$) diagram. The Bp stars, pulsating stars and normal 
B stars are represented by triangles, circles and squares, respectively. The full lines represent boundaries of 
theoretical instability strips for modes with frequency between 0.25 and 25 c d$^{-1}$ and $\ell \leq 4$, computed for main sequence models with 2 M$_\odot$ $\leq M \leq$ 15 M$_\odot$ by De Cat et al.\ (\cite{decat2}). 
The lower and upper dotted lines show the ZAMS and TAMS, respectively. The dashed lines denote evolution tracks 
for stars with $M = 12, 9, 6,$ and 3\,M$_\odot$. Filled symbols correspond to stars with detected magnetic fields.}
\label{HR}
\end{figure*}

Because the colours of Bp stars are anomalous, the usual photometric calibrations allowing to estimate $\log g$ for normal B stars (e.g. in the Geneva system) are much less reliable for them; photometry can still provide useful information in a statistical sense, but only for very large samples. Few $\log g$ values determined spectroscopically are available, especially for SPB stars. Therefore, the only 
consistent way to determine the position of the stars of the three samples in the H-R diagram is to use Hipparcos 
parallaxes. For this study we selected exclusively stars for which very accurate parallaxes, 
(i.e., with $\sigma (\pi)/ \pi < 0.2$) and Geneva or Str\"omgren photometry are available. Our whole sample 
consists of 24 Bp stars, 24 SPB stars and 5 normal B stars. 
They are listed in Table\,\ref{Bp_SPB_B}.

\section{Determination of stellar parameters}

The effective temperature was determined using reddening-free photometric parameters in the Geneva photometric system through the calibration of K\"unzli et al.\ (\cite{kunzli}). In the case of Bp stars, the calibration was corrected for the anomalous colours according to the prescriptions described by North\ (\cite{north2}). The luminosity was obtained from Hipparcos parallaxes using bolometric corrections measured by Lanz\ (\cite{lanz}). 

Following the recommendation by Arenou\ (\cite{arenou}) that the Lutz-Kelker correction (Lutz \& Kelker\ \cite{lutz}) should not
be used for individual stars, no such correction was applied to the absolute magnitudes of our targets.

For binary systems, a duplicity correction to the magnitude was taken into account. For SB2 systems, this correction is estimated from mass 
ratios available in the literature. For SB1 systems a statistical correction of 0.1 mag was adopted. In our SPB sample, 
HD\,123515,  HD\,140873, HD\,24587, HD\,53921, HD\,74560, HD\,92287 and HD\,160762 are SB systems according to 
De Cat et al.\ (\cite{decat3}), De Cat \& Aerts\ (\cite{decat_aerts}) and Abt \& Levi\ (\cite{abt}). It means that $\sim$ 1/3 of SPB stars in our sample belong 
to binary systems. This number is representative for SPB stars in general (De Cat\ \cite{decat1}). In our Bp sample, there is only 
one SB1 system: HD\,25823 (Wolff\ \cite{wolff}). The lack of binary systems among Bp stars had already been noticed by 
Gerbaldi et al.\ (\cite{gerbaldi}).

The interstellar reddening was taken into account for stars farther away than 100 pc, from reddening-free parameters 
X and Y of Geneva photometry and Cramer's intrinsic [UBV] colours (Cramer\ \cite{cramer}) and verified using the interstellar absorption maps of Lucke\ (\cite{lucke}). The latter precaution was needed for a few stars, the reddening of which is overestimated by Cramer's method, due to the anomalous reddening law in the region of the Upper Sco association.

The mass was obtained from interpolation in the evolutionary tracks of Schaller et al.\ (\cite{schaller}) assuming
solar metallicity. It is generally expected that Bp stars follow standard, solar composition evolutionary tracks 
and the surface chemical
anomalies are produced by the process of selective radiative diffusion in the presence of a magnetic field.
The radius was directly computed  from luminosity and effective temperature with 

\smallskip

$\log (R/R_\odot) = \frac{1}{2} \log(L/L_\odot) - 2 \log(T_{\rm eff}/T_{\rm eff \odot})$. 

\smallskip

Finally, the surface gravity was obtained from mass and radius through its fundamental definition. 
More details on the determination of the stellar parameters can be found in Hubrig et al.\ (\cite{hubrig1}).
  
The basic data for the three samples, Bp stars, SPB stars and normal B stars, are 
presented in Table\,\ref{Bp_SPB_B}. Note that, for HD\,125823, we considered the more accurate effective temperature 
and gravity derived by Hunger \& Groote\ (\cite{hunger}) by means of the IR flux method.
The distribution of all targets in a ($\log T_{\rm eff},\log g$) diagram is shown in Fig.~\ref{HR}. 

\section{Comparisons}
\subsection{Evolutionary state}
\begin{figure}
\centering
\includegraphics[width=0.35\textwidth]{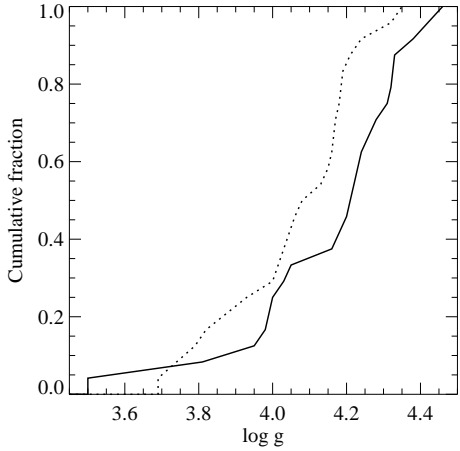}
\caption{Cumulative distribution of $\log g$ for the Bp stars (full line) and the SPB stars (dotted line).}
\label{cumu}
\end{figure}

It is quite clear from Fig.~\ref{HR} that the majority of Bp stars is rather young, with a location close to 
the zero-age main sequence (ZAMS). The SPB stars are significantly older whereas normal B-type stars are distributed 
over the whole width of the main sequence. 
In Fig.~\ref{cumu} we show the cumulative distribution of $\log g$ for the studied Bp and SPB stars. We note that 
the $\log g$ range of SPB stars perfectly agrees with evolutionary main-sequence models; the same is true of Bp stars, if one excepts HD 5737, which has the smallest surface gravity of all stars of its kind in our sample. But this object also has the largest relative error on its parallax, and its $\log g$ value is less than 2$\sigma$ smaller than the minimum value of the SPB stars. A Kolmogorov-Smirnov test shows that the distribution of $\log g$ values for the Bp stars differs from the distribution 
for SPB stars at a significance level of 98.3 \%. Obviously, the group of Bp stars is younger than the group of SPB stars. The conclusion remains valid if we compare distributions of radius rather than $\log g$.
It is of further interest that the position of SPB stars with detected magnetic fields indicates that they are 
younger than SPB stars with no magnetic detections. However, to confirm this clue, further systematic searches for magnetic fields 
in a larger sample of SPB stars should be conducted.

\begin{figure}
\centering
\includegraphics[width=0.35\textwidth]{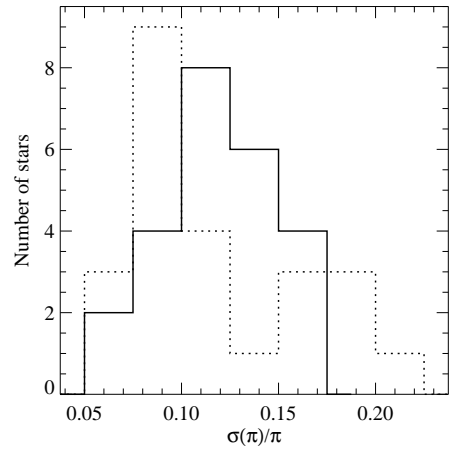}
\caption{Distribution of the relative errors of the Hipparcos parallaxes for the Bp stars (full line) and the SPB stars (dotted line).}
\label{spi}
\end{figure}

As a check that this conclusion is not biased by the distributions of the relative errors of Hipparcos parallaxes we displayed this 
distribution in Fig.~\ref{spi} as a histogram for both samples. The average parallax uncertainty is 12 \% for both Bp and SPB stars. 
According to a Kolmogorov-Smirnov test, the distributions do not differ at a significant level, so that no systematic bias has been 
introduced in the $\log g$ cumulative distributions. 

\begin{figure}
\centering
\includegraphics[width=0.35\textwidth]{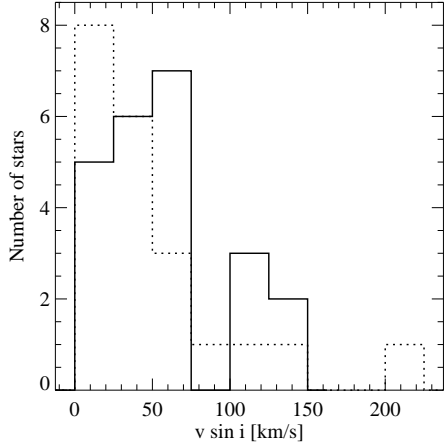}
\caption{Distribution of $v$\,sin\,$i$ values for the Bp stars (full line) and the SPB stars (dotted line) in our sample.}
\label{vsini}
\end{figure}

\begin{figure}
\centering
\includegraphics[width=0.35\textwidth]{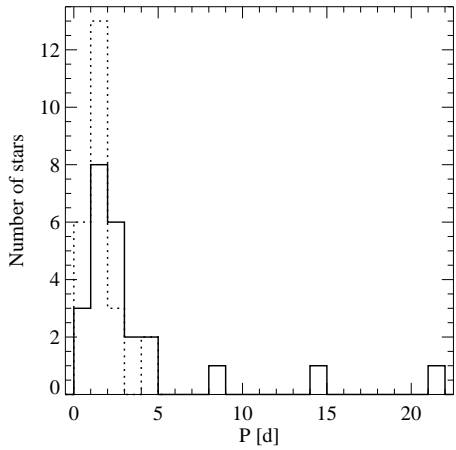}
\caption{Distribution of rotation periods $P$ of Bp stars (full line) and of pulsation periods of SPB stars 
(dotted line).}
\label{p}
\end{figure}

\begin{figure}
\centering
\includegraphics[width=0.35\textwidth]{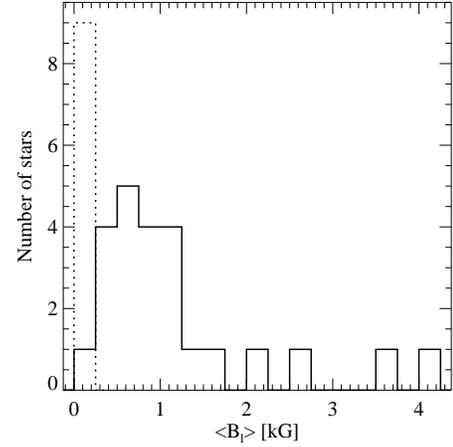}
\caption{Distribution of the longitudinal magnetic field values $\left<B_l\right>$ for the Bp stars (full line) and the SPB stars (dotted line).}
\label{mag}
\end{figure}

\begin{figure}
\centering
\includegraphics[width=0.35\textwidth]{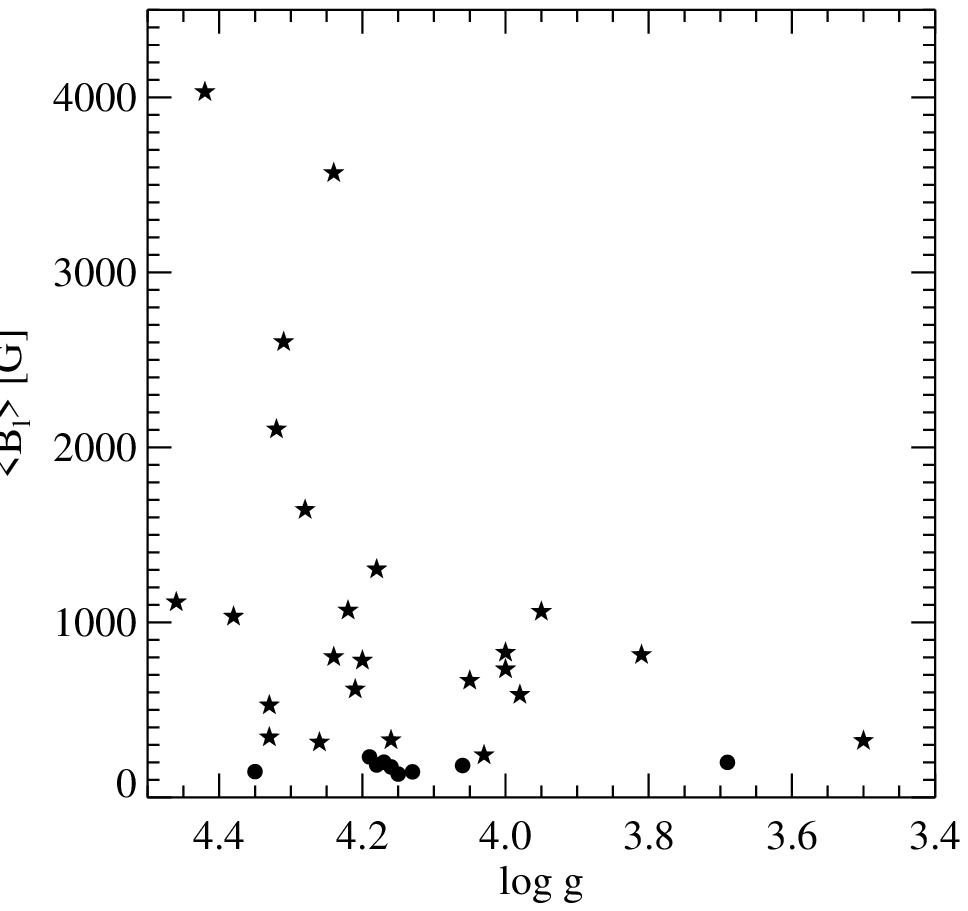}
\caption{Averaged quadratic effective magnetic field for Bp stars (filled stars) and SPB stars (filled circles)
versus $\log g$.
}
\label{logg_Heff}
\end{figure}

\subsection{Rotation and magnetic field strength}

It is well-known that both Bp and SPB stars are slow rotators. The distributions of the projected rotational 
velocities $v$\,sin\,$i$ acquired from the SIMBAD data base for the sample of Bp stars and of SPB stars are shown in Fig.~\ref{vsini}.
Apparently, they are slightly different with a maximum in the bin 0--25\,km/s for SPB stars and a maximum 
around 50--75 \,km/s for Bp stars. 

The knowledge of rotation periods of SPB stars is very poor and only very few of them 
have periods indirectly determined by mode identification methods (De Cat et~al.\ \cite{decat4}). On the other hand, it is remarkable that the pulsation periods of these stars are very similar 
to the rotation periods of Bp stars. This fact led to some confusion in previous studies of SPB stars, where the variations of 
spectral line profiles 
with an inhomogeneous distribution of Si and He had first been assumed to be caused by non-radial pulsations (Briquet et~al.\ \cite{briquet}).
In  Fig.~\ref{p} we present the distributions of the rotation periods of Bp stars and pulsation periods of SPB stars. The periods, given in Table\,\ref{Bp_SPB_B} in column 2, were retrieved from 
the catalogue of magnetic rotational phase curves of CP stars by Bychkov et al.\ (\cite{bychkov2}) and from De Cat\ (\cite{decat1}). Both distributions are very similar in that they show a maximum between 1 and 2 days.

Magnetic Bp stars generally have large-scale organized magnetic 
fields. Most studies of their magnetic fields are based on measurements of the
mean longitudinal magnetic field which is an average over the visible stellar
hemisphere of the component of the magnetic vector along the line of sight.
Bychkov et al.\ (\cite{bychkov1}) presented a catalogue of averaged stellar effective magnetic fields of chemically peculiar A and B type stars. For the SPB stars in our sample 
we used the rms mean longitudinal magnetic fields determined by Hubrig et al.\ (\cite{hubrig3}). The rms mean longitudinal magnetic fields are presented in column 3 of Table\,\ref{Bp_SPB_B}.
The distribution presented in Fig.~\ref{mag} clearly shows 
that longitudinal magnetic fields in SPB stars are significantly weaker in comparison to the magnetic fields 
detected in Bp stars. 
In Fig.~\ref{logg_Heff} we present the  evolution of the averaged quadratic effective magnetic 
field $\left<B_l\right>$ in Bp and SPB stars over the main sequence. 
The value $\log g$ is used as a proxy for the relative age and has the advantage
of being a directly measured quantity.
It is quite obvious that the strongest magnetic fields appear in young Bp stars.
The fact that strong magnetic fields are only observed in a restricted 
range of evolutionary states can be interpreted as a hint for a magnetic field decay in stars at advanced ages.
A similar result has already been presented by Hubrig et~al.\ (\cite{hubrig5})
from magnetic field measurements with FORS\,1 at the VLT.

It is especially intriguing that the
magnetic fields of a few Bp stars either do not show any detectable variations or
vary with periods close to one day, which is of the order of the pulsation
period range of SPB stars
(Bohlender et al.\ \cite{Bohlender1987}, Matthews \& Bohlender\ \cite{Matthews1991}). 
The study of stellar parameters of the He-strong star 
HD\,96446 with a rotation period of 0.85\,d by Mathys\ (\cite{Mathys1994}) revealed that the radius determined from 
considerations of the observed magnetic field structure is far too 
small and does not correspond to the spectral type of this star. He suggested that pulsation could be a possible 
candidate to explain the variation of the 
magnetic field. 
Another example suggests that pulsation might be the cause of magnetic field variability in some B-type stars.
Bychkov et~al.\ (\cite{bychkov2}) showed that the detected magnetic field of the star HD\,37151 varies with a period
whose one-day-alias is interpreted as a pulsation period by North \& Paltani\ (\cite{north}),
who also proved this star to be a multiperiodic SPB star. However, further magnetic measurements are necessary to confirm this clue because the rotational phase curve presented for the star only relies on few magnetic data with large error bars. 

\section{Geneva photometry}

\begin{figure}
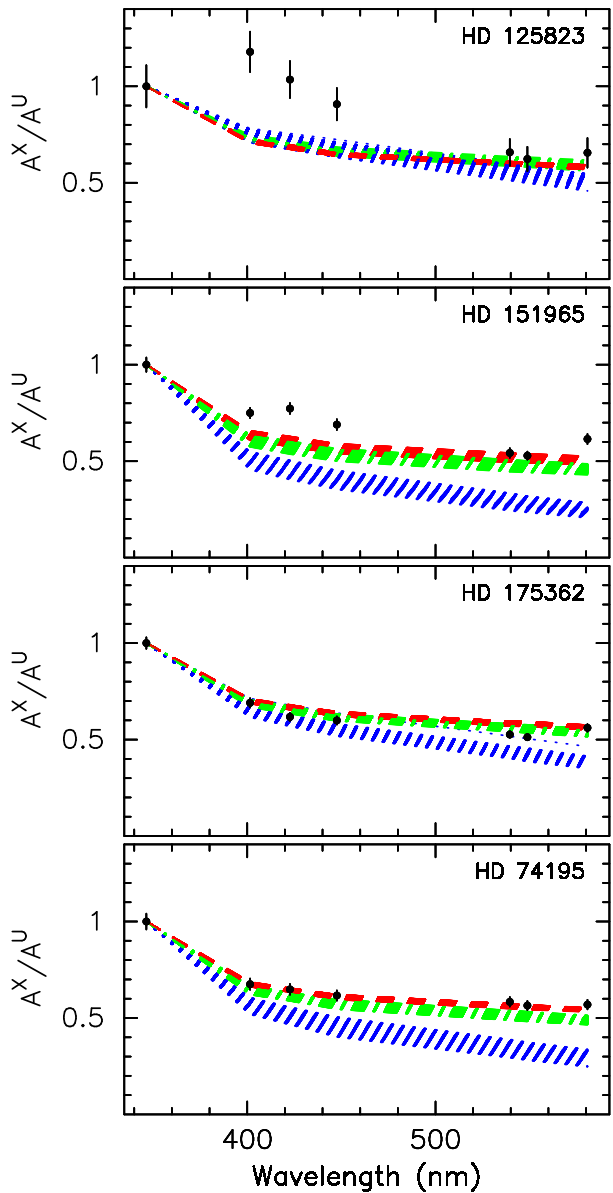

\centering
\includegraphics[angle=-90,width=8cm]{6940fig8_1.ps}\\
\includegraphics[angle=-90,width=8cm]{6940fig8_2.ps}\\
\includegraphics[angle=-90,width=8cm]{6940fig8_3.ps}\\
\includegraphics[angle=-90,width=8cm]{6940fig8_4.ps}\\
\caption{
Comparison between observed photometric amplitude ratios and the amplitudes predicted by stellar pulsation theory.
The theoretical amplitude ratios for modes with $\ell$\,= 1, 2, 3, and 4 are represented
with a red dashed, a green dash-dotted, a blue dotted, and a cyan dash-dot-dot-dotted line, respectively
(the colour representation is only available in the online version of the paper).
The dots indicate the observed amplitude ratios and their standard error.
}
\label{MI}
\end{figure}

\begin{figure}
\centering
\includegraphics[width=8.2cm]{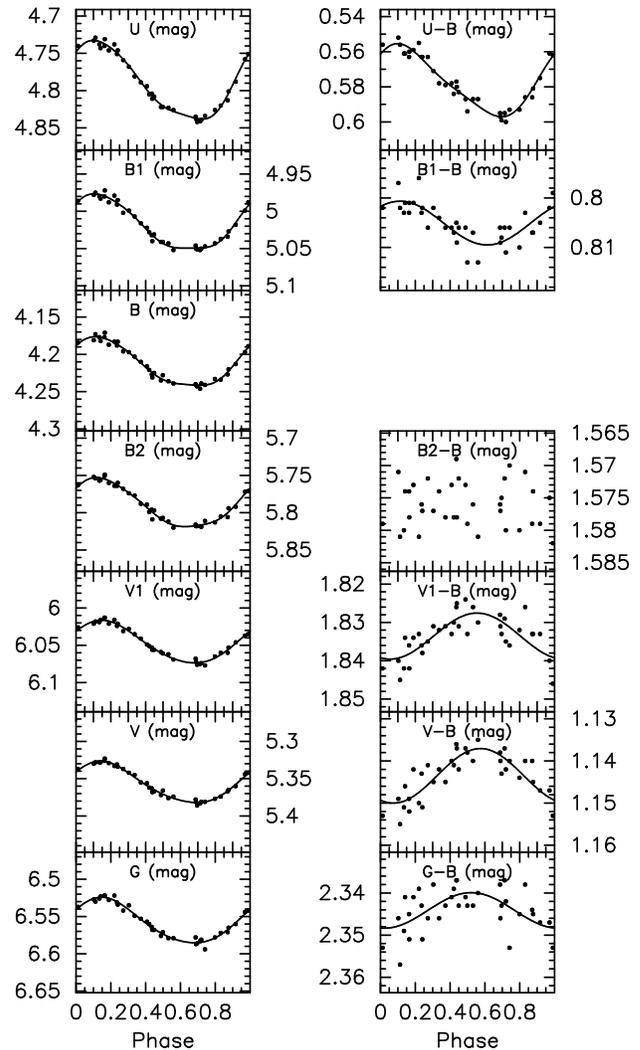}
\caption{Geneva photometric variability of HD\,175362. }
\label{hd175362}
\end{figure}

As already mentioned above, the discrimination between the pulsation and rotation modulation
interpretations is not obvious.
With the aim to find the cause of the observed variability of B-type stars,
we made a comparison of Geneva photometry for our sample of Bp stars and SPB stars
to study the behaviour of passband and colour variability.
In particular, we computed the amplitude ratios,
commonly used to identify pulsation modes (e.g., Dupret et al.\ \cite{dupret}). 

The characteristics of the Geneva data for SPB stars can be summarized as follows.
The data in the seven Geneva filters vary as a sine function for each oscillation frequency and are all in phase.
The amplitude is the highest in the U-filter and the typical behaviour of the amplitude ratios computed
against the U-filter is shown in Fig.\,\ref{MI} for the main mode observed for HD\,74195.
In respect to colours, the variation is dominant in U-B and only stars with the highest amplitudes show
clear variability in the other colours, which are thus all in phase or antiphase.

The types of photometric variability of Bp stars are more diversified than for SPB stars (e.g., North\ \cite{north0}).
In almost all cases, the data are only fitted well when one considers the frequency and at least one of its harmonics, which makes a first difference compared to the SPB stars. The amplitude is not necessarily the highest in the U-filter but many of them have a larger amplitude in the B-filter. Variations in colours are generally much more pronounced for Bp stars than for SPB stars and they are not always in phase or antiphase.

For the Bp stars for which we have sufficient Geneva data at our disposal,
we computed amplitude ratios in order to compare them with typical cases of SPB stars.
We found two kinds of behaviour, different from what is predicted by pulsation theory.
The first one is represented for the star HD\,125823 (a Centauri) in Fig.\,\ref{MI}.
This star is one of the best studied among hot peculiar stars
for which the changes in the line strength of the He lines is so conspicuous
that the star can be considered as a He-weak star at one phase and a He-rich star at another phase (Norris\,1968).
Such a behaviour of the photometric amplitude ratios as well as very large equivalent width variations
were also observed for the star HD\,55522 (Briquet et al.\ \cite{briquet}).
The other kind of observed amplitude ratio behaviour is shown in Fig.\,\ref{MI}
and is present in HD\,151965.
In this case, the amplitudes for the B1-, B-, and B2-filters are smaller than that for the U-filter,
but remain too large to be fully compatible with a pulsation model.
For some stars, the amplitude in the V-filter was found also to be too small (Briquet et~al.\ \cite{briquet2}).

There is also another feature found in the Bp stars studied, which concerns their strict monoperiodicity.
When the main period with its harmonics is removed from the data set,
the residual standard deviations are of the order of the error on the data,
indicating that no additional variability is present
or that the variability is of much smaller amplitude than the pulsation amplitudes of SPB stars.

Several Bp stars also show amplitude ratios indistinguishable from the ones of SPB stars.
In Fig.\,\ref{MI} the amplitude ratios of the star HD\,175362 are displayed. The observed ratios are fully in agreement with the theoretical amplitude ratios predicted by pulsation theory. For details on how theoretical amplitude ratios for a pulsating star are computed, we refer to De Cat et~al.\ (\cite{decat2}). The variability in the Geneva passbands and colours are presented in Fig.\,\ref{hd175362}.
Apart from the presence of the first harmonics in the data and an amplitude in the U-filter ($\sim$54 mmag)
larger than the one for typical SPB stars (De Cat\ \cite{decat1}),
the behaviour of the Geneva photometry is similar to the one for an SPB star.
In the literature, the variablitity of HD\,175362 is interpreted in terms of the oblique rotator model.
For instance, Hensler\ (\cite{hensler}) modelled the star with a single He-cap and a single Si-cap located at the opposite magnetic poles.    

The case of HD\,175362 illustrates that a spotted star might have photometric variability in perfect agreement with a pulsation model. In that case, it might be that periodicity interpreted as pulsation is actually due to rotation and only spectroscopy can help to make the differentiation.
However, one can alternatively explain the variability of HD\,175362 by non-linear stellar pulsation instead of rotational modulation.
Indeed, there is a striking analogy to the $\beta$~Cephei star $\xi^1$~CMa,
for which the pulsation interpretation is without doubt.
The passbands, colour and amplitude ratio behaviour of HD\,175362 is completely similar to the one of $\xi^1$~CMa,
also in relation to the presence of harmonics in the data set and the large amplitude of the variability.
$\xi^1$~CMa pulsates non-linearly and is apparently monoperiodic.
If additonal modes are present in this pulsating star, they are of much smaller amplitudes than the main mode (Saesen et al.\ \cite{saesen}).     
Interestingly, we recently discovered a magnetic field in $\xi^1$~CMa of the order of 300\,G (Hubrig et al.\ \cite{hubrig3}).
 
From a literature search, we summarize as follows the characteristics of spectral lines of HD\,175362
and their phase relation with the photometric and magnetic field data.
Balona\ (\cite{balona}), Wolff \& Wolff\ (\cite{wolff_wolff}), Hensler\ (\cite{hensler}) and Catalano \& Leone\ (\cite{catalano})
report that the Si\,II line strengths vary in antiphase with helium,
that the velocity variations are in quadrature with the variations of line strength,
and that the light curves are in antiphase with respect to the He line strength.
This does not seem to be incompatible with a pulsation model (De Ridder\ \cite{deridder}). However, a thorough and quantitative comparison is necessary to definitely support a pulsation model or not. This is beyond the scope of this paper. 
Wolff \& Wolff\ (\cite{wolff_wolff}), Borra et al.\ (\cite{borra}), Bohlender et al.\ (\cite{Bohlender1987}),
Mathys\ (\cite{mathys}) and Mathys \& Hubrig\ (\cite{mathys_hubrig}) obtained magnetic field measurements and
detected for HD\,175362 a strong magnetic field with a non-sinusoidal variation (see Table\,\ref{Bp_SPB_B}).
The light extrema coincide with the magnetic field extrema
and spectral variations of several elements (Si, C, Fe, Ga) vary in phase with the magnetic field.
It remains to be shown that pulsation may produce the observations reported by the authors above and cause the magnetic variability (Mathys\,\cite{mathys99}).
We found no argument to definitely favour one model.

\section{Discussion}

Our study of the evolutionary age of magnetic Bp and SPB stars with accurate Hipparcos parallaxes and available 
Geneva photometry revealed a clear difference in their ages at the significance level of 98.3\%.
The Bp stars show much stronger magnetic fields than the SPB stars and are younger as a group. An interesting possibility raised by these results is that at least some Bp stars may transform themselves into SPB stars as they become older.
Only one Bp star in our sample belongs to an SB1 system whereas $\sim$1/3 of the SPB stars are members of SB systems.
Unfortunately, we could not make a statistical comparison of the distribution of our Bp and SPB star samples with
that for normal B-type stars as only five such stars have accurate Hipparcos parallaxes and available Geneva or 
Str\"omgren photometry. 

Variation periods of Bp and SPB stars are of the same order.
Such similar distributions of periods with a maximum between 1 and 2~days
makes the interpretation of the observed variability of B-type stars located in the instability strips of SPB stars quite hard.
The difficulty is increased by the fact that stellar rotation with spots
and stellar pulsation may lead to very similar behaviour of the observed photometric variability. The example of HD 175362 teaches us, that a Bp star which actually is a spotted and oblique rotator could well be interpreted in terms of a pulsation model; but that conversely, it is not excluded that a star actually pulsating non-linearly with a dominant main mode and showing a large and strongly variable magnetic field, could be wrongly modelled in terms of an oblique rotator with abundance spots.

The magnetic field variability of Bp stars is generally interpreted in terms of the oblique rotator model.
However, the cases of HD\,96446 and HD\,37151 suggest that pulsation might also be the cause for magnetic field variability.
Clearly, additional magnetic field measurements of SPB stars are needed in order to search for possible relations between the magnetic and pulsation variability and between the field strength and the amplitudes of the pulsation modes. The failure to find multiperiodic signals in Bp stars indicate that in B-type stars very strong magnetic fields are not coexistent with oscillations, or stars with stronger magnetic fields have much lower pulsation amplitudes. 

Both Bp and SPB stars are slow rotators.
A slow rotation rate is consequently not the only condition for a star to become chemically peculiar.
As already suggested by Michaud\ (\cite{michaud}), the magnetic field stabilizes the atmosphere,
permitting diffusion processes to become important.
The magnetic field strength is very likely an important factor in explaining why some stars
are chemically peculiar while others are not. Besides, one may suspect that pulsations of the SPB kind tend to inhibit radiative diffusion, which would explain why SPB stars are not chemically peculiar in spite of the small magnetic field they sometimes exhibit. We finally point out that the strongest magnetic fields appear in young Bp stars, indicating a magnetic field decay in main-sequence stars at advanced ages.    

The evolutionary state of magnetic chemically peculiar stars has already been studied
by several authors in the literature with the aim to better understand the origin of magnetic fields in Ap/Bp stars (e.g., Hubrig et al.\ \cite{hubrig1,hubrig5}).
Recently, Kochukhov \& Bagnulo\ (\cite{kochukhov}) found that magnetic stars with $M > 3M_\odot$ are homogeneously distributed along the main sequence.
Our study based on a smaller, yet carefully selected, sample of stars,
showed that the majority of Bp stars are rather young stars with a location close to the zero-age main sequence (ZAMS).
In any case, the observation of young magnetic stars is in favour of the fossil theory.
Direct confirmation of the presence of magnetic fields in the pre-main sequence phase was
recently provided by the discovery of magnetic fields in Herbig Ae/Be stars (Hubrig et al.\ \cite{hubrig6,hubrig7,hubrig8}, Wade et al.\ \cite{wade}),
which are considered as the progenitors of main sequence early-type stars.

%\begin{acknowledgements}
%\end{acknowledgements}

{}

\end{document}